\documentstyle[aps,prb]{revtex}
\begin{document}

\input epsf.sty
\twocolumn[\hsize\textwidth\columnwidth\hsize\csname %
@twocolumnfalse\endcsname

\draft

\widetext

\title{Static and dynamic spin correlations in the spin-glass phase
of slightly-doped La$_{2-x}$Sr$_x$CuO$_4$}
\author{M. Matsuda$^*$}
\address{
RIKEN (The Institute of Physical and Chemical Research),
Wako, Saitama 351-0198, Japan}
\author{M. Fujita and K. Yamada}
\address{
Institute for Chemical Research, Kyoto University, Gokasho, Uji
610-0011, Japan}
\author{R. J. Birgeneau and M. A. Kastner}
\address{
Department of Physics and Center for Materials Science and Engineering,
Massachusetts Institute of Technology, Cambridge,
Massachusetts 02139}
\author{H. Hiraka and Y. Endoh}
\address{
Institute for Materials Research, Katahira, Sendai 980-8577, Japan}
\author{S. Wakimoto$^\dagger$ and G. Shirane}
\address{
Department of Physics, Brookhaven National Laboratory, Upton, New
York 11973}

\date{28 June, 2000}
\maketitle
\begin{abstract}
Neutron scattering experiments reveal that
a diagonal spin modulation, which is a one-dimensional
modulation rotated away by 45$^\circ$ from that in the superconducting
phase, occurs universally across the insulating spin-glass phase
in La$_{2-x}$Sr$_x$CuO$_4$ (0.02$\le x\le$0.055).
This establishes an intimate relation between the magnetism
and the transport properties in the high-temperature copper oxide
superconductors.
Furthermore, it is found that the charge density per unit length
estimated using a charge stripe model is almost constant throughout
the phase diagram, even when the modulation rotates away by 45$^\circ$
at the superconducting boundary. However, at the lowest values
for $x$ the density changes approaching 1 hole/Cu as in
La$_{2-x}$Sr$_x$NiO$_4$.
Magnetic excitation spectra suggest that magnetic correlations change
from incommensurate to commensurate at $\omega\sim$7 meV and
$T\sim$70 K, indicating a characteristic energy for the
incommensurate structure of 6-7 meV.

\end{abstract}

\phantom{.}
]
\narrowtext

\section{Introduction}
The phase diagram of La$_{2-x}$Sr$_x$CuO$_4$ shows that the magnetic state
changes dramatically with Sr doping. The parent material La$_2$CuO$_4$
exhibits three-dimensional (3D) long-range antiferromagnetic (AF) order
below $\sim$325 K. \cite{lco} When a small fraction of La is replaced
by Sr, which corresponds to hole-doping, the 3D AF order disappears
and the low temperature magnetic phase is replaced by a disordered
magnetic phase in which commensurate two-dimensional (2D) short-range AF
fluctuations are observed. \cite{sternlieb,keimer} In crucible-grown
samples the commensurate fluctuations develop a static component at low
temperatures, signalling the onset of spin-glass order.

In superconducting samples, an essential feature is that the magnetic
correlations become incommensurate (IC).
\cite{cheong,yamada,katsumata,birgeneau}
Detailed studies on the hole concentration dependence of the low
energy magnetic
excitations have been performed by Yamada $et$ $al.$ \cite{yamada} They
find that the incommensurability ($\delta$) is almost linear with
hole concentration ($x$) with $\delta\simeq x$ below $x\sim$0.12.
Recently, static magnetic ordering has been observed in superconducting
La$_{1.88}$Sr$_{0.12}$CuO$_4$ \cite{suzuki,kimura}
with the magnetic onset temperature near $T_c$.
The elastic magnetic peaks are observed at the same IC positions as
those of the magnetic inelastic peaks. A model that describes this
behavior is that of stripe ordering of spin and charge (hole) density
waves as observed in La$_{2-y-x}$Nd$_y$Sr$_x$CuO$_4$.
\cite{tranquada,tranquada2} In this case the charge
and, concomitantly, spin stripes run approximately along the
$a\rm_{tetra}$ or $b\rm_{tetra}$ axis; we label this the collinear
stripe phase.

Thus, the magnetism and the transport properties in the doped
La$_2$CuO$_4$ system are intimately related. \cite{kastner}
In the insulating phase at low hole concentrations,
spin-glass behavior is observed and there are strong quasi-elastic
commensurate spin fluctuations; dynamic IC spin fluctuations
persist in the superconducting phase. It has been known for some time that
in La$_{2-x}$Sr$_x$CuO$_4$ the instantaneous magnetic correlations change
from being commensurate to IC at the
insulator-to-superconductor boundary.
Recently, Wakimoto $et$ $al.$ have found that in a sample grown with the
crucible-free traveling solvent floating zone technique which results
in purer crystals the static magnetic correlations at low temperature
are also IC in the insulating spin-glass La$_{1.95}$Sr$_{0.05}$CuO$_4$.
\cite{wakimoto} They have examined the intensity profiles and have
shown that there are only 2 satellite peaks
along $b\rm_{ortho}$ \cite{wakimoto2} while in superconducting compounds
the IC peaks are located parallel to both the $a\rm_{tetra}$ and
$b\rm_{tetra}$ axes. These magnetic correlations
in La$_{1.95}$Sr$_{0.05}$CuO$_4$
are consistent with diagonal charge stripes, in which the stripes run along
the $a\rm_{ortho}$ axis. Actually, such diagonal stripes have been
predicted theoretically. \cite{machida0,kato,rice,schulz,zaanen}
Diagonal stripes are also reported experimentally in insulating
La$_{2-x}$Sr$_x$NiO$_4$. \cite{tranquada3}
We emphasize, however, that only a
one-dimensional \underline{spin} modulation has been observed
in La$_{2-x}$Sr$_x$CuO$_4$ to-date;
any associated \underline{charge} ordering has not yet been detected.
These results lead to the important
conclusion that the static magnetic spin modulation changes from diagonal
to collinear at $x=0.055\pm0.005$, coincident with the
insulator-to-superconductor transition.

A fundamental question is whether or not the diagonal one-dimensional
IC magnetic correlations persist throughout the spin-glass phase down to
the critical concentration of $x$=0.02 for 3D N\'{e}el ordering.
The present neutron scattering study clarifies this point and
yields important new information on the concentration dependence of
the incommensurability. Especially, we find that at the lowest
concentration within the context of the stripe model the inferred charge
density is $\sim$1 hole/Cu as in La$_{2-x}$Sr$_x$NiO$_4$.

Another important point is to clarify the nature of the magnetic
excitations in the diagonal IC state.
Intensive studies of the inelastic magnetic spectra in insulating
La$_{2-x}$Sr$_x$CuO$_4$ were performed by Keimer $et$ $al.$ \cite{keimer}
They studied the energy and temperature dependences of the $Q$-integrated
susceptibility. However, the $Q$-dependence of the excitation spectra was
not discussed since the detailed peak profile was not known.
Matsuda $et$ $al.$ also studied the inelastic magnetic spectra in
La$_{1.98}$Sr$_{0.02}$CuO$_4$, \cite{matsuda} in which the
energy and temperature dependences of the excitation spectra were
measured. They only discussed the results qualitatively since
the detailed peak profile could not be clarified.
Now that the static magnetic correlations have been elucidated, the
excitation spectra can be analyzed qualitatively. Specifically,
we found that the magnetic correlations change from being incommensurate
to commensurate at $\omega\sim$7 meV and $T\sim$70 K,
indicating a characteristic energy for the IC structure of 6-7 meV.

\section{Experimental details}
The single crystal of La$_{1.976}$Sr$_{0.024}$CuO$_4$ was grown by the
traveling solvent floating zone (TSFZ) method. The crystal was
annealed in an Ar atmosphere at 900 $^\circ$C for 24 h.
The dimensions of the rod shaped crystal were
$\sim$5$\Phi\times$25 mm$^{3}$.
The lattice constants were $a\rm_{ortho}$=5.349 \AA,
$b\rm_{ortho}$=5.430 \AA\ ($b/a\sim$1.015), and $c$=13.151 \AA\ at 10 K.
From the universal relation for the spin-glass transition temperature,
the tetragonal-to-orthorhombic structural transition temperature,
and the orthorombicity $b/a$,\cite{yamada3} the effective hole
concentration was estimated to be 0.024$\pm$0.003.

The neutron scattering experiments were carried out on the
cold neutron three-axis spectrometer HER and the thermal neutron
three-axis spectrometer TOPAN installed at JRR-3M at the Japan Atomic Energy
Research Institute (JAERI). The horizontal collimator sequences were
guide-open-S-80$'$-80$'$ with the fixed incident neutron energy $E\rm_i$=5 meV
at HER and 30$'$-30$'$-S-60$'$-60$'$ with the fixed final neutron energy
$E\rm_f$=14.7 meV at TOPAN. Contamination from higher-order beams was
effectively eliminated using Be filters at HER and PG filters at TOPAN.
The single crystal, which was oriented in the $(HK0)\rm_{ortho}$ or
$(H0L)\rm_{ortho}$ scattering plane, was mounted in a closed cycle
refrigerator. In this paper, we use the low temperature
orthorhombic phase ($Bmab$) notation $(h,k,l)\rm_{ortho}$
to express Miller indices.

The crystal has a twin structure and there exist two domains. The
two domains are estimated to be equally distributed from the ratio
of the nuclear Bragg peak intensities from both domains.
Figure 1A shows the scattering geometry in the $(HK0)$ scattering
plane. The filled triangles correspond to the (1,0,0) and (0,1,0) Bragg
points from domain A while the open triangles denote the (1,0,0) and
(0,1,0) Bragg points from domain B.

\section{Results and Discussion}
\subsection{Static properties}
Below $\sim$40 K elastic magnetic peaks develop and
at low temperatures the peaks are clearly resolved at the IC
positions (1, $\pm\epsilon$, 0) and (0, 1$\pm\epsilon$, 0)
with $\epsilon\sim$0.023. This corresponds to the same
diagonal one-dimensional spin modulation observed in
La$_{1.95}$Sr$_{0.05}$CuO$_4$ which has $\epsilon\sim$0.064.
\cite{wakimoto} The open and filled circles in Fig. 1A correspond
to the IC magnetic peaks from the two domains in the $(HK0)$ zone,
respectively.

Figures 1B-1D show transverse and longitudinal elastic scans
around (1,0,0) and (0,1,0). Two peaks are observed in the transverse
scan A while one intense peak together with a weak shoulder on the
low-$h$ side is observed in the longitudinal scans B and C.
The instrumental resolution at (1,0,0) can be estimated from
higher-order reflections, which in turn are measured by removing the Be
filters. As illustrated in Figs. 1B and 1C, the magnetic peaks
are much broader than the resolution along both $h$ and $k$.
It should be noted that, by contrast, the magnetic peaks
in the superconducting state of La$_{1.88}$Sr$_{0.12}$CuO$_4$
are all resolution limited, \cite{kimura} indicating that the magnetic
correlation length in that system is quite large in the CuO$_2$ plane.

Figure 2 shows the $L$-dependence of the magnetic elastic
peaks at (1,0,$L$) and (0,0.975,$L$), respectively, at 10 K.
The background estimated from the high temperature data (100 K) has
been subtracted so that the remaining signal is purely magnetic.
These scans probe the magnetic correlations along the $c$ axis
between neighboring 2D antiferromagnetically correlated planes.
Broad peaks are observed at $(1,0,even)$ and (0,$\sim$1,$odd$) which
coincide with the magnetic Bragg peak positions in pure La$_2$CuO$_4$.
\cite{lco} However, the magnetic intensity at $(1,0,even)$ initially
increases with increasing $L$, implying a cluster spin-glass model
as in La$_{1.98}$Sr$_{0.02}$CuO$_4$: \cite{matsuda}
the spin system forms antiferromagnetically correlated clusters which have
randomly different spin directions in the CuO$_2$ plane although the
propagation vector of the AF order is along $a\rm_{ortho}$ in each cluster.
\begin{figure}
\centerline{\epsfxsize=3in\epsfbox{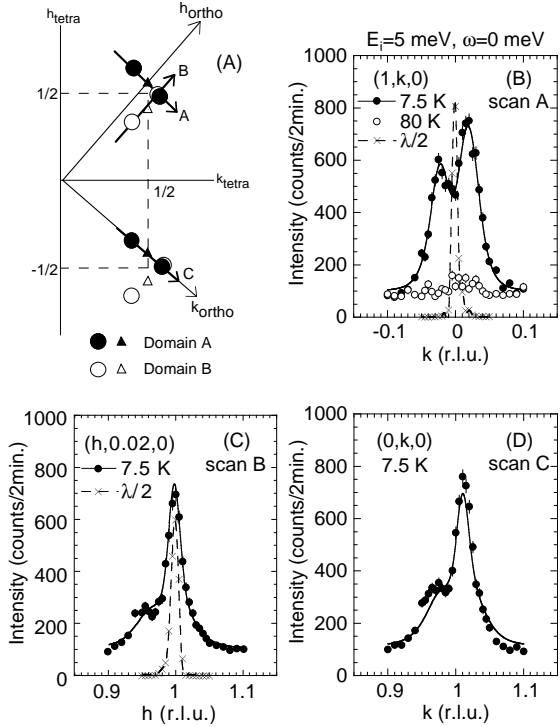}}
\caption{(A) Diagram of the reciprocal lattice in the $(HK0)$
scattering zone. Filled and open symbols are for domains A and B,
respectively. The triangles and circles correspond to nuclear and
magnetic Bragg peaks, respectively. The thick arrows show scan
trajectories.
Transverse (B) and longitudinal elastic scans (C) and
(D) around (1,0,0) and (0,1,0) at 7.5 K (filled circles) and 80 K
(open circles).
The crosses represent the higher-order Bragg peaks observed at (1,0,0)
by removing the Be filters. The broken lines are guides to the eyes.
The peak width represents the instrumental resolution. The solid
lines are the results of fits to a convolution of the resolution
function with 3D squared Lorentzians with $\xi'_a$=94.9 \AA,
$\xi'_b$=39.9 \AA, $\xi'_c$=3.15 \AA, and $\epsilon$=0.0232 r.l.u.}
\label{fig1}
\end{figure}

The solid lines in Figs. 1B-1D are the results of fits to a
convolution of the resolution function with 3D squared Lorentzians.
The two intense peaks in Fig. 1B originate primarily from the
magnetic signals at (1,$\pm\epsilon$,0) in domain A while the weak
shoulder in Fig. 1C originates from magnetic signals at
(0,1$-\epsilon$,$\pm$1) in domain B. The relatively intense peaks
at (0,1$\pm\epsilon$,0) occur because of the short correlation
length along the $c$ axis, which in turn makes the
(0,1$\pm\epsilon$,$L$), with $L$ odd, magnetic peaks broad along
the $c$ axis as shown in Fig. 2.
The instrumental resolution function is also elongated along the
$c$ axis so that the magnetic signals are effectively integrated.
The observed data are fitted with $\xi'_a$=94.9$\pm$4.0 \AA,
$\xi'_b$=39.9$\pm$1.3 \AA, $\xi'_c$=3.15$\pm$0.08 \AA, and
$\epsilon$=0.0232$\pm$0.0004 r.l.u., where $\xi'_{a}$, $\xi'_{b}$,
and $\xi'_{c}$ represent the inverse of the half width at half maxima
of the elastic peak widths in $Q$ along
the $a$, $b$, and $c$ axes, respectively.
The calculation reproduces the observed profiles quite well.
The error bars represent one standard deviation statistical error
limits for the assumed lineshape. The true error limits, indicating
possible systematic errors are much larger.
The static correlation length perpendicular to the CuO$_2$ plane
is 3.15 \AA, which is much less than the distance
between nearest-neighbor CuO$_2$ planes ($c/2\sim$6.5 \AA), indicating
that the static magnetic correlations are almost two-dimensional.
The between-plane correlation length is shorter than those
in La$_{1.48}$Nd$_{0.4}$Sr$_{0.12}$CuO$_4$ ($\xi_c \sim 0.55c$ at 1.38 K)
\cite{tranquada2} and La$_{1.775}$Sr$_{0.225}$NiO$_4$
($\xi_c \sim 1.06c$ at 10 K). \cite{tranquada3}

The solid lines in Fig. 2 show the calculated profiles using
3D squared Lorentzian profiles convoluted with the instrumental
resolution function. The parameters determined above
are held fixed and only the overall scale factor has been adjusted.
In order to reproduce the $L$-dependence of the $(1,0,even)$ intensity,
the cluster spin-glass model, \cite{matsuda} as described above,
has been used in the calculation. The calculation
describes the observed profiles reasonably well. The slight deviation at
large $L$ in both the $(1,0,L)$ and $(0,0.975,L)$ scans probably
reflects a decrease at large $L$ of the magnetic form factor, which has
been assumed to be constant in the calculation.
\begin{figure}
\centerline{\epsfxsize=3in\epsfbox{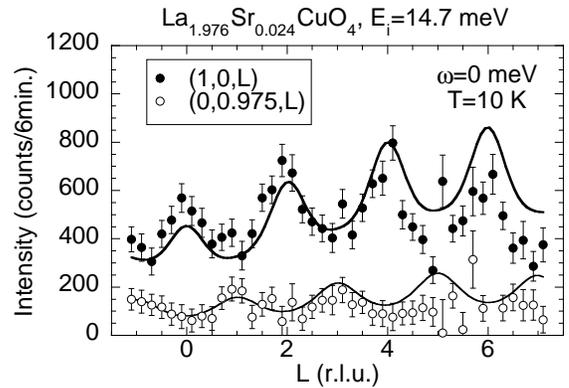}}
\caption{Elastic scans along $(1,0,L)$ and along (0,0.975,$L$) at
10 K. The background intensities measured at 100 K have been subtracted.
The solid lines show the results of calculations with $\xi'_a$=94.9 \AA,
$\xi'_b$=39.9 \AA, $\xi'_c$=3.15 \AA, and $\epsilon$=0.0232 r.l.u.}
\label{fig2}
\end{figure}

The diagonal magnetic stripe model thus provides a good description
of the data; this is one of the most significant results of this study.
The incommensurability $\epsilon$ corresponds to the
inverse modulation period of the spin density wave. Here, $\epsilon$
is defined in orthorhombic notation so that
$\epsilon=\sqrt{2}\times\delta$ where $\delta$ is defined in
tetragonal units. As shown in Fig. 3, $\delta$ follows the linear
relation $\delta=x$ reasonably well over the range 0.03$\le x\le$0.12
which spans the insulator-superconductor transition. In a charge
stripe model this corresponds to a constant charge per unit length in
both the diagonal and collinear stripe phases, or equivalently, 0.7 and
0.5 holes per Cu respectively because of the $\sqrt{2}$ difference in
Cu spacings in the diagonal and collinear geometries. Our value for
$x$=0.024 definitely deviates from the $\delta =x$ line and instead
appears to be close to $\sim$1 hole/Cu as in La$_{2-x}$Sr$_x$NiO$_4$ where
there is $\sim$1 hole/Ni \cite{tranquada3} along the diagonal stripes.
This suggests that as the hole concentration is decreased, in the
context of the stripe model, the hole concentration evolves
progressively from $\sim$0.5 hole/Cu at $x$=0.12 to 1 hole/Cu
at $x$=0.024.
This behavior is very different from that in La$_{2-x}$Sr$_x$NiO$_4$,
where the hole density is $\sim$1 hole/Ni over a wide
range of hole concentrations in the insulating phase albeit at rather
larger hole densities. \cite{yoshizawa}
We note that Machida and Ichioka predict 1 hole/Cu throughout the
diagonal stripe phase. \cite{machida}
\begin{figure}
\centerline{\epsfxsize=3in\epsfbox{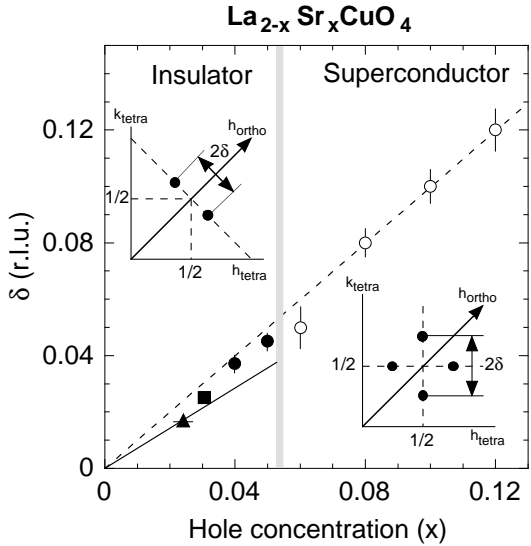}}
\caption{Hole concentration ($x$) dependence of the splitting of
the IC peaks ($\delta$) in tetragonal reciprocal lattice units.
Open circles indicate the data for the
inelastic IC peaks reported by Yamada $et$ $al.$
(5). Filled circles and square are the data for the
elastic IC peaks reported by Wakimoto $et$ $al.$
(14,27). The filled triangle is obtained from the present
study. The broken and solid lines correspond to $\delta =x$ and
$\epsilon =x$, respectively. The insets
show the configuration of the IC peaks in the insulating phase
(diagonal stripe) and the superconducting phase (collinear stripe).}
\label{fig3}
\end{figure}

The static spin correlation lengths, which are derived from the inverse
peak widths in $Q$, in La$_{2-x}$Sr$_x$CuO$_4$ ($x$=0.02, 0.024,
and 0.05) are summarized in Table 1.
With increasing hole concentration, the peak widths both parallel
and perpendicular to the CuO$_2$ plane rapidly broaden.
In La$_{1.95}$Sr$_{0.05}$CuO$_4$, an $L$-scan shows that the peak
width along $L$ is much broader than that in
La$_{1.976}$Sr$_{0.024}$CuO$_4$, indicating that the magnetic
correlations are more two-dimensional.~\cite{wakimoto2}

There are at least two possible origins of the finite correlation
lengths in the CuO$_2$ plane for the static order in the spin-glass
La$_{2-x}$Sr$_x$CuO$_4$.
The first is that the lengths simply measure the spin decoherence
distance of the AF spin clusters. The second is that the disorder
originates primarily from a random distribution of stripe spacings and
orientations as discussed by Tranquada $et$ $al$.~\cite{tranquada4}
It is noted that a rotation of the stripe orientation away from the
$a\rm_{tetra}$ and $b\rm_{tetra}$ axes is observed in
La$_2$CuO$_{4+y}$ (Ref. \onlinecite{lee}) and La$_{1.88}$Sr$_{0.12}$CuO$_4$
(Ref. \onlinecite{kimura2}).
Further experiments and theoretical calculations will be required to
choose between these possibilities.

\subsection{Magnetic excitations}
As mentioned in the previous section, static magnetic correlations in
the spin-glass phase has been revealed.
We now consider how the inelastic magnetic correlations behave in the
spin-glass phase of La$_{2-x}$Sr$_x$CuO$_4$.
Neutron inelastic scattering measurements were performed in
La$_{1.976}$Sr$_{0.024}$CuO$_4$. The measurements were performed in
the ($H0L$) scattering plane. We first carried out the measurements in
the ($HK0$) scattering plane, in which the elastic IC magnetic
peaks can be resolved reasonably well since the instrumental resolution is
narrower than the intrinsic peak widths as shown in Fig. 4(a).
However, with increasing transfer energy, the resolution becomes worse
and the magnetic intensity decreases due to the structure factor.
Because of these two effects, it is very difficult to follow the energy
dependence of the excitation spectra.
In the ($H0L$) scattering plane, since the resolution is elongated
perpendicular to $H$, as shown in Fig. 4(b), the IC peak
cannot be well resolved. However, as we will show later, the
intrinsic peak configuration can be estimated by fitting to model
functions. The advantage of the measurements in the ($H0L$) scattering
plane is that the measurements can be performed with increased scattering
intensity.

\begin{table}
\caption{Hole concentration dependence of the static spin correlation
length in La$_{2-x}$Sr$_x$CuO$_4$. $-$ means that the peak width is
too broad to determine $\xi'_c$. It is noted that $\xi'_b$ in the
$x$=0.02 sample is obtained on the assumption that the magnetic
correlations are commensurate. This value could become larger
($\sim$45 \AA) if the magnetic correlations are incommensurate with
$\epsilon=x$.}
\label{table1}
\begin{tabular}{cccc}
$x$ & $\xi'_a$ (\AA) & $\xi'_b$ (\AA) & $\xi'_c$ (\AA)\\ \tableline
0.02\tablenote{Ref. 21.}& 160 & 25 & 4.7\\
0.024\tablenote{This work.}& 94.9$\pm$4.0 & 39.9$\pm$1.3 & 3.15$\pm$0.08\\
0.05\tablenote{Ref. 14.}& 25 & 33 & $-$\\
\end{tabular}
\end{table}
\begin{figure}
\centerline{\epsfxsize=3in\epsfbox{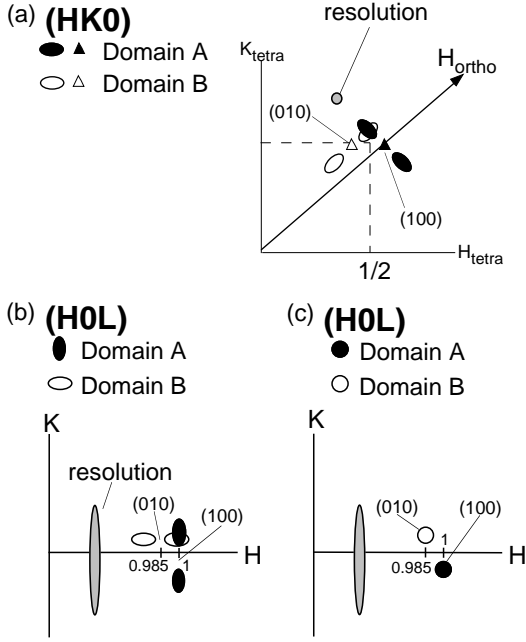}}
\caption{The schematic configuration of the magnetic peaks
in the $(HK0)$ and $(H0L)$ scattering planes.
The instrumental resolution is elongated perpendicular to the
scattering plane. The ellipsoids and circles represent magnetic peaks.
(a) and (b) show scattering configurations in the IC phase.
As shown in the text, the IC magnetic peaks are anisotropic.
In the commensurate phase (c), the peaks are considered to be isotropic.}
\label{fig4}
\end{figure}
First, we performed constant-$Q$ scans around ($\pi,\pi$) in order
to study the magnetic anisotropy in La$_{1.976}$Sr$_{0.024}$CuO$_4$.
Figure 5(a) shows the result of constant-$Q$ scans at (0.992,0,0.7),
which corresponds to just the midpoint between (1,0,0.7) and (0,1,0.7).
The background intensities are measured at (1.15,0,0.7). The magnetic
intensity decreases gradually with increasing energy, indicating that the
magnetic excitation spectrum is gapless in La$_{1.976}$Sr$_{0.024}$CuO$_4$.
For comparison, constant-$Q$ scans in pure La$_2$CuO$_4$ with
the same spectrometer condition are shown in Fig. 5(b).
An excitation gap due to the out-of-plane anisotropy is found at $\sim$5 meV,
which is consistent with that observed previously. \cite{keimer2}

Figure 6 shows constant-$\omega$ scans in the ($H0L$) scattering
plane at various energies and temperatures. A sharp excitation peak is
centered at $H$=1 at $\omega$=3 meV and $T$=10 K. On the other hand, the peak
position shifts progressively to lower $H$ at higher energies and
temperatures.

\begin{figure}
\centerline{\epsfxsize=3in\epsfbox{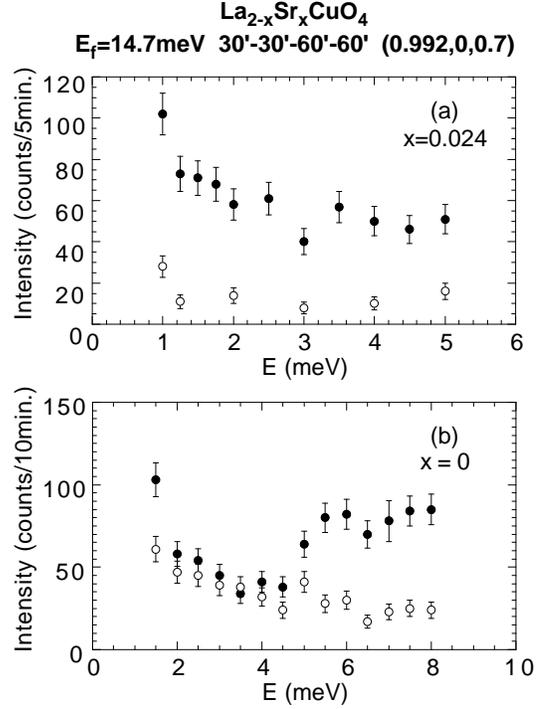}}
\caption{Filled circles show constant-$Q$ scans at (0.992,0,0.7)
measured at $T$=10 K in La$_{1.976}$Sr$_{0.024}$CuO$_4$ (a) and
in La$_2$CuO$_4$ (b). Open circles show background intensities
measured at (1.15,0,0.7).}
\label{fig5}
\end{figure}
We speculate that this behavior may be explained as follows.
At 3 meV, the magnetic peaks exist at IC positions as observed in the
elastic scans in the ($HK0$) scattering zone as shown in Fig. 4(a).
Since the instrumental resolution elongated vertically integrates
the magnetic signal around $H$=1 very effectively in the $(H0L)$
scattering plane, as shown in Fig. 4(b), a sharp and intense peak
centered at $H$=1 is observed while a weak tail is found at lower $H$.
The solid line in the
Fig. 6(a) is the result of a calculation assuming that the magnetic
peaks are located at exactly the same positions as those determined
from the elastic measurements in the ($HK0$) scattering plane.
As excitation energy is increased, the peak separation appears to
to become smaller. The excitation spectrum at 6 meV is fitted
to 3D Lorentzians. The fitting parameters are the
peak separation $\epsilon$, the isotropic inverse inelastic peak width
in the CuO$_2$ plane $\xi''_{ab}$, and the amplitude.
The inverse inelastic
peak width along the c-axis $\xi''_c$ is fixed at 3.15 \AA, which is
the same as $\xi'_c$. The solid line in the Fig. 6(b) is the result of a
fit to the 3D Lorentzians with
$\epsilon$= 0.0096$\pm$0.0068 r.l.u. and $\xi''_{ab}$=48$\pm$12 \AA.
Finally, the magnetic correlations appears to become commensurate
and isotropic at 9 meV as shown in Fig. 4(c).
In this case, there exist two equi-intense peaks at (0,1,-0.6)
($H$=0.985) and (1,0,-0.6) ($H$=1) resulting in one broad peak located
at $H\sim$0.99. The solid line in Fig. 6(c) is the result of a
fit to the 3D Lorentzians with $\xi''_c$=3.15 \AA\ (fixed)
and $\xi''_{ab}$=42$\pm$13 \AA. $\epsilon$ is fixed at 0 r.l.u.
since it becomes very close to 0 r.l.u. even when fitted.
From these results, we conclude that magnetic correlations change
from being incommensurate to commensurate between $\sim$6 and
$\sim$7.5 meV.
\begin{figure}
\centerline{\epsfxsize=3in\epsfbox{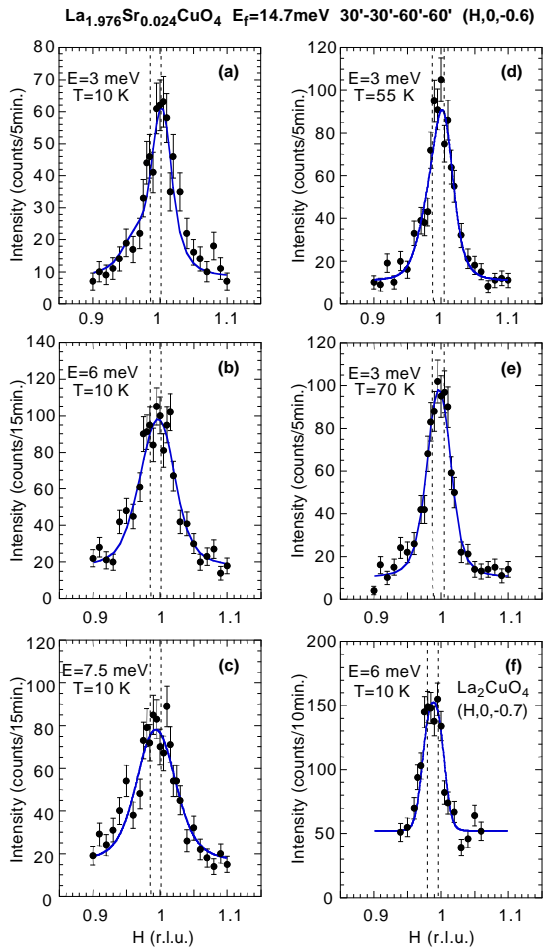}}
\caption{Neutron inelastic scans along $(H,0,-0.6)$ and $(H,0,-0.7)$ at various
energies and temperatures in La$_{1.976}$Sr$_{0.024}$CuO$_4$ and in
La$_2$CuO$_4$. The solid lines are the results of fits to a convolution
of the resolution function with 3D squared Lorentzians.
The broken lines show the centers of the peaks (1,0,$L$) and (0,1,$L$)
determined from the nuclear Bragg peak positions.}
\label{fig6}
\end{figure}

The commensurate magnetic correlations at higher energies are
similar to those observed in pure La$_2$CuO$_4$ above the
gap energy $\sim$5 meV as shown above. The excitation spectrum
in pure La$_2$CuO$_4$ at 6 meV is shown in Fig. 6(f).
The spectrum is consistent with the peak configuration as shown in
Fig. 4(c) since the magnetic correlations are commensurate and
isotropic. The peak width is resolution-limited, indicating that the
correlation length is very long in the CuO$_2$ plane.
This is in striking contrast to the situation in
La$_{1.976}$Sr$_{0.024}$CuO$_4$.
The solid line in Fig. 6(f) is the result of a calculation
assuming the 3D Lorentzians with $\xi''_{ab}$=700 \AA\ (fixed),
$\xi''_c$=1 \AA\ (fixed), and $\epsilon$= 0 r.l.u. (fixed).

The temperature dependence of the excitation spectra in
La$_{1.976}$Sr$_{0.024}$CuO$_4$ is quite similar
to the energy dependence.
A sharp excitation peak centered at $H$=1 at low temperatures
shifts to lower $H$ with increasing temperature. 
The solid line in the Fig. 6(d) is the result of a fit to an assumed
3D Lorentzian line-shape with $\epsilon$= 0.0113$\pm$0.0068
r.l.u. and $\xi''_{ab}$=60$\pm$11 \AA.  $\xi''_c$ is fixed at 1 \AA\ 
since the system becomes magnetically 2D above $\sim$40 K.
$\epsilon$ and $\xi''_{ab}$ do not change sensitively with changes in
$\xi''_c$ in the fitting.
The solid line in Fig. 6(e) is the result of a
fit to a 3D Lorentzian with $\xi''_{ab}$=62$\pm$10 \AA\ and
$\xi''_c$=1 \AA\ (fixed). $\epsilon$ is fixed at 0 r.l.u. since
it becomes very close to 0 r.l.u. even when fitted.
From these results, we conclude that the magnetic correlations change
from being incommensurate to commensurate between 55 and 70 K.

\begin{figure}
\centerline{\epsfxsize=3in\epsfbox{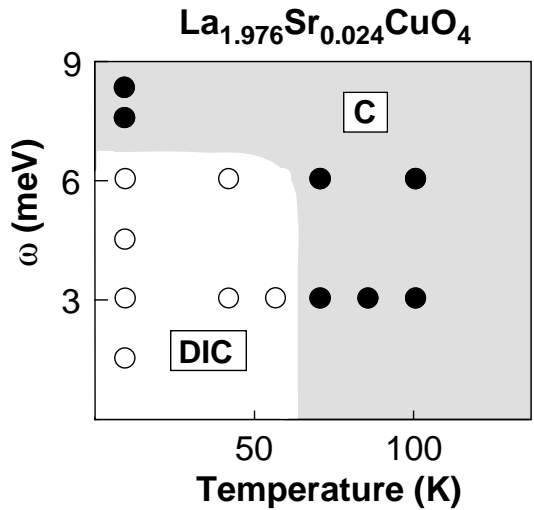}}
\caption{Energy-temperature phase diagram in
La$_{1.976}$Sr$_{0.024}$CuO$_4$. Open and filed circles represent that
the magnetic correlations are diagonal incommensurate (DIC) and
commensurate (C), respectively.}
\label{fig7}
\end{figure}
Figure 7 represents a summary of the neutron inelastic measurements in
La$_{1.976}$Sr$_{0.024}$CuO$_4$. The open and filled circles
signify that the magnetic correlations are diagonal IC
and commensurate, respectively. The diagonal IC phase
exists below $\omega\sim$7 meV and $T\sim$70 K ($\sim$6 meV). This
result indicates that the characteristic energy for the diagonal
IC structure is 6-7 meV.

\section{Summary}
In brief, we find that a short range static one-dimensional diagonal
spin modulation exists at low temperatures across the entire insulating
spin-glass region in La$_{2-x}$Sr$_x$CuO$_4$. Further, within the
context of a spin and charge stripe model the charge
density per unit length is almost constant for all values of $x$,
but shows a significant deviation near the spin-glass 3D N\'{e}el
boundary suggesting stability of diagonal stripes with 1 hole/Cu
at low $x$.

The magnetic excitation spectra suggest that magnetic correlations
change from diagonal incommensurate to commensurate at
$\omega\sim$7 meV and $T\sim$70 K. Above these energy and temperature
the magnetic correlations are similar to those in pure La$_2$CuO$_4$
although the range of order in the CuO$_2$ plane is much shorter
in La$_{1.976}$Sr$_{0.024}$CuO$_4$.

\section*{Acknowledgments}
We would like to thank A. Aharony, K. Katsumata, and K. Machida for
stimulating discussions. This study was supported in part by the
U.S.-Japan Cooperative Program on Neutron Scattering, by a Grant-in-Aid
for Scientific Research from the Japanese Ministry of Education, Science,
Sports and Culture, by a Grant for the Promotion of Science from the
Science and Technology Agency, and by CREST.
Work at Brookhaven National Laboratory was carried out under Contract
No. DE-AC02-98CH10886, Division of Material Science, U.S. Department of
Energy. The research at MIT was supported by the National Science
Foundation under Grant No. DMR97-04532 and by the MRSEC Program of
the National Science Foundation under Award No. DMR98-08941.

\end{document}